\documentclass[a4paper]{jpconf}
\usepackage{subcaption}
\usepackage{graphicx}
\usepackage{amsmath,amssymb,multirow,bm}
\usepackage[colorlinks,linkcolor=blue,anchorcolor=blue,citecolor=blue,urlcolor=blue,filecolor=black]{hyperref}

\newcommand{\be}{\begin{equation}}
\newcommand{\ee}{\end{equation}}
\newcommand{\ba}{\begin{eqnarray}}
\newcommand{\ea}{\end{eqnarray}}

\begin{document}

\title{Cluster Structures with Machine Learning Support in Neutron
  Star M-R relations}
\author{Ronaldo V. Lobato$^{1,3}$, Emanuel V. Chimanski$^{2}$ and
  Carlos A. Bertulani$^{3}$}

\address{$^{1}$Departamento de F\'isica, Universidad de los Andes, Bogot\'a, Colombia.\\
  $^{2}$Lawrence Livermore National Laboratory, Livermore, CA, USA.\\
  $^{3}$Department of Physics and Astronomy, Texas A\&M University - Commerce,
  Commerce, TX, USA.}

\ead{r.vieira@uniandes.edu.co, chimanski1@llnl.gov, carlos.bertulani@tamuc.edu}

\begin{abstract}
Neutron stars (NS) are compact objects with strong gravitational fields, and a matter composition subject to extreme physical conditions. The properties of strongly interacting matter at ultra-high densities and temperatures impose a big challenge to our understanding and modelling tools. Some difficulties are critical, since one cannot reproduce such
conditions in our laboratories or assess them purely from astronomical observations. The information we have about neutron star interiors are often extracted indirectly, e.g., from the star mass-radius relation. The
mass and radius are global quantities and still have a significant uncertainty, which leads to great variability in studying the micro-physics of the neutron star interior. This leaves open many questions in nuclear astrophysics and the suitable equation of state (EoS) of NS. Recently, new observations appear to constrain the mass-radius and consequently has helped to close some open questions. In this work, utilizing modern machine learning techniques, we
analyze the NS mass-radius (M-R) relationship for a set of EoS containing a variety of physical models. Our objective is to determine patterns through the M-R data analysis and develop tools to understand the EoS of neutron stars in forthcoming works.
\end{abstract}

\section{Introduction}
Neutron stars (NS) are one of the densest objects in the Universe, and the geometry of the spacetime around it deviates considerably from flat spacetime. These
objects have densities in the range of few ${\rm g/cm^3}$ to more than $10^{15}\ {\rm g/cm^3}$ in their centers \cite{haensel/2007}. Such extreme density and temperature conditions are expected to affect the properties of interacting matter. The microscopic description, i.e, the equation of state (EoS) of NS have been extensively studied. In recent years, new data from gravitational waves, optical, X- and gamma-rays
and from satellites in other bandwidths, composed a new research area of multi-messenger astronomy and some joint constraints have been achieved
\cite{margalit/2017, radice/2018a, tews/2018, motta/2019, gamba/2019, lourenco/2020}. These efforts have helped to shade light on the path to the EoS.

Recently, the current authors have explored \cite{lobato/2022}
correlations among a small set of equation of states derived from different models, where the sample studied was the most common utilized in the literature
\cite{lattimer/2001, lackey/2006, bejger/2005, ozel/2016}. We have studied correlations among the many parameters that generate those EoS, i.e.,
the microscopic description of each EoS. In \cite{lobato/2022} we limited our analysis to just one global property, the maximum mass reached for each EoS. In this work we are going to
extend the set utilized, studying global properties such as
mass, radius, maximum mass and correlations with the properties of the equations of state. In Section \ref{sec:mr} we
give a description of the mass-radius relationship and its
connection with the EoS, i.e., the connection between the micro and macro physics. In Section \ref{sec:cluster} we discuss the machine-learning clustering method employed. Finally, in Section \ref{sec:conc} we give our finals remarks.

\section{From micro to macroscopic properties}\label{sec:mr}
The micro-physics approach to the EoS comes from the quantum theory of many-body systems. It depends on the highly complex theory of super-dense matter, and on the degrees of freedom of QCD. The connection with
global properties comes through the Einstein's field equation, i.e.,
the models of neutron stars need to be constructed in the framework of
Einstein's general relativity equations,
\begin{equation}
  G^{\mu\nu} \equiv R^{\mu\nu} - \frac{1}{2}g^{\mu\nu}R = 8\pi
  T^{\mu\nu}(\rho, P(\rho)),
  \end{equation}
where the matter field is coupled to the gravitational field by the energy-momentum tensor. This tensor,
$T^{\mu\nu}$, contains the equation of state $P(\rho)$. The
many-body equations generating the EoS, are solved in flat
spacetime, entering as {\it a posteriori} input for the Einstein's field
equations. Therefore, to associate the microscopic properties' connection to the astronomical
data, one needs to readjust the quantum theories' parameters, recalculate the many-body equations and regenerate
the global properties in a process of backwards feedback since the gravitation
field equations and the many-body equations are not solved
simultaneously. One can use the machine learning technique to inverse the approach, such as in inverse problems \cite{glasko/1984,prilepko/2000, samarskii/2008, aster/2018} and, starting with the global properties, find the underlying features of the dense matter, as proposed, e.g., in Ref.~\cite{fujimoto/2018}.

To model neutron stars one needs to use the theory of general relativity and find out the
hydrostatic equilibrium equation, reduced to the Tolman-Oppenheimer-Volkoff (TOV)
equation \cite{tolman/1939, oppenheimer/1939} describing a static spherical neutron star. The hydrostatic equilibrium equations read in natural units
\begin{equation}\label{tov}
  p' = - (\rho + p) \frac{4\pi pr + m/r^2}{(1 - 2m/r)},
\end{equation}
where prime indicates radial derivative, $r$ is the radial coordinate, $p$ the pressure, $\rho$ the energy density, and $m$ is the gravitational mass enclosed within the surface of radius $r$, i.e., $m' = 4\pi\rho r^2.$

To solve the TOV equation, one needs to provide the EoS with the boundaries conditions $m(0)=0$, $p(0)=p_c$
and $\rho(0) = \rho_c$ at the center of the star ($r=0$). Propagating the solution to $r=R$, the stellar
surface, the pressure must vanish there, i.e., $p(R)=0$. The total gravitational mass is obtained from
\begin{equation}
M(R) \equiv 4 \pi \int_0^R r^2 \rho(r) dr.
\end{equation}
The M-R relation outputs are what we call global properties, and they can be compared to astronomical data. For each EoS, an M-R relation is obtained. One can construct a bilinear map between the EoS space ($V$) and the M-R relation space ($W$). In \cite{lobato/2022} we have
studied correlations in the former space, while here we are
interested in correlations of the latter. One can go further and study the correlations in the $V\otimes W$ space.

One can compare the
correlations in $V$ with the vicinity of the nuclear saturation
density, which recently started to gather valuable data \cite{reed/2021}; for
$W$, one can compare with the gravitational waves
\cite{ligo/2018, ligo/2019} and the {\it NICER} \cite{miller/2019a, riley/2019} constraints for the mass-radius relation.

\begin{figure}[h]
  \centering

  \begin{subfigure}{0.75\textwidth}
    \includegraphics[width=\textwidth]{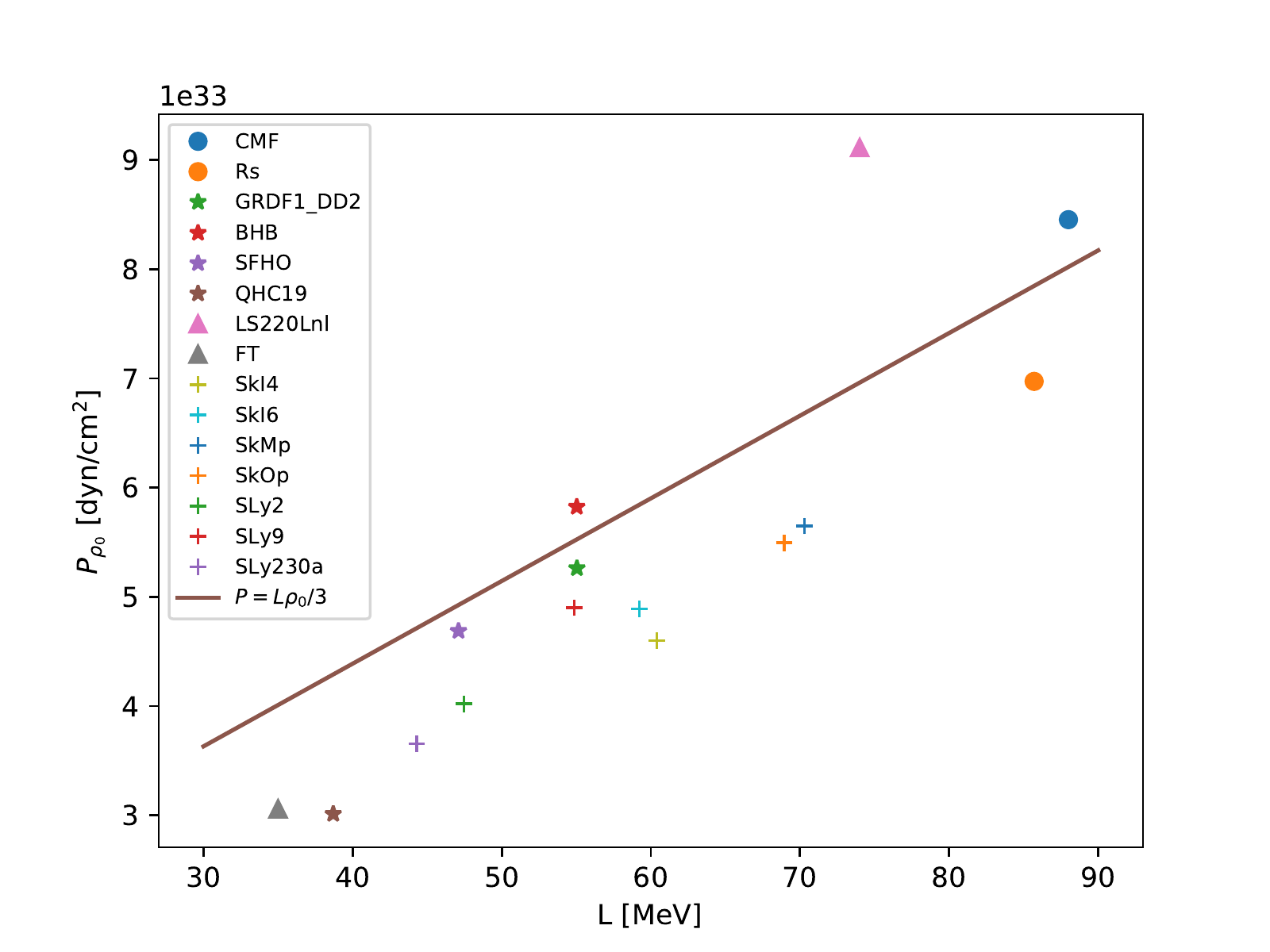}
    \caption{Pressure in the neighborhood of the nuclear saturation density vs
      the slope parameter of the symmetry energy for a set of EoS in colored
      geometric shapes. The brown
      line represents Eq. \eqref{slope}.}
    \label{p2L}
  \end{subfigure}

  \begin{subfigure}{0.75\textwidth}
    \includegraphics[width=\textwidth]{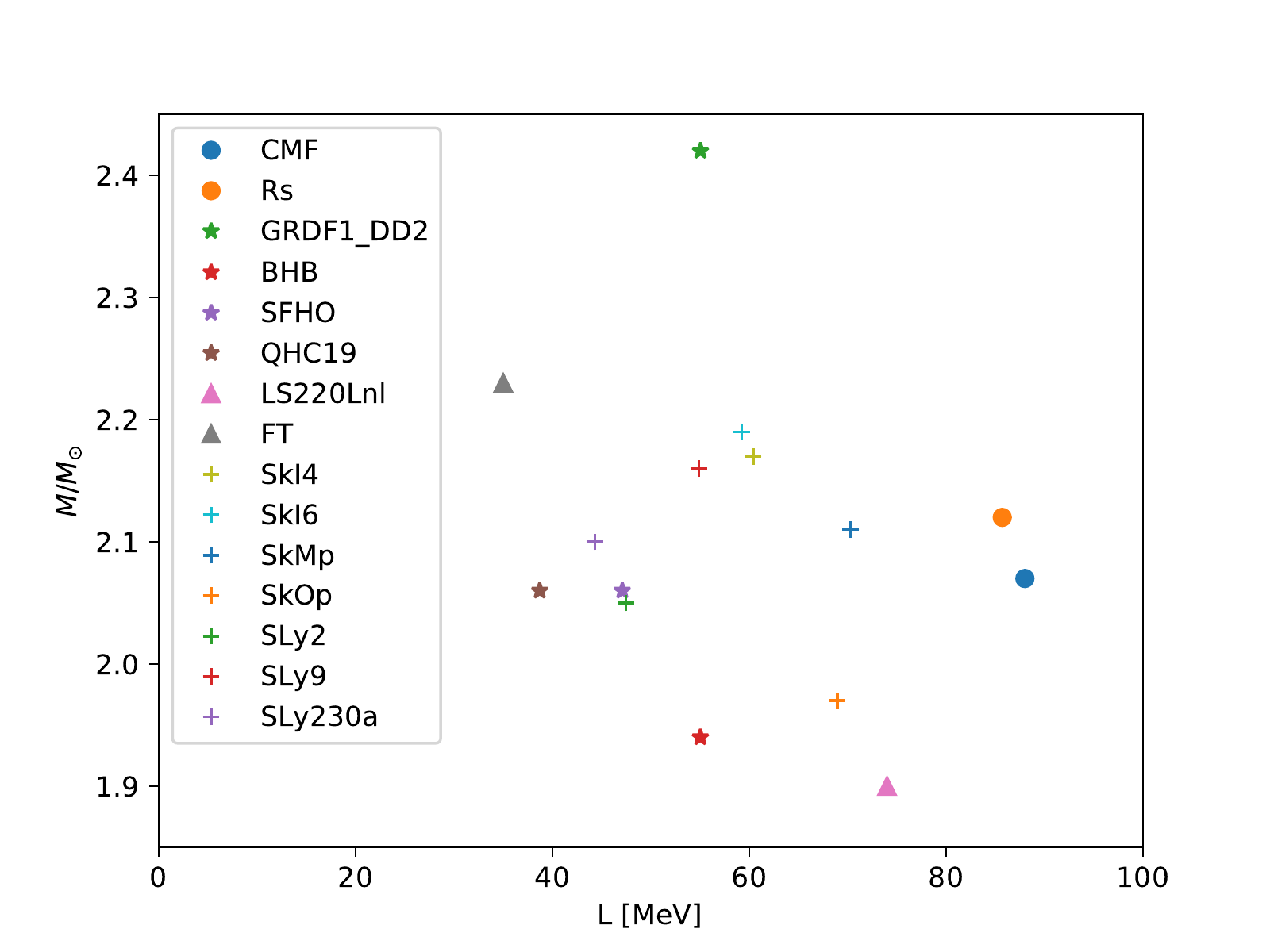}
    \caption{Maximum mass vs the slope parameter of the symmetry energy for
    a set of EoS in colored geometric shapes.}
    \label{mlall}
  \end{subfigure}

  \caption{Pressure and maximum mass vs the slope parameter of the symmetry energy.}
\end{figure}

In Fig. \ref{p2L} we show a set of EoS for an energy range 30--90 MeV of the slope
of the symmetry energy. For symmetric nuclear matter, the slope of the
symmetry energy, $L$, is closely related to the pressure by means of
\begin{equation}\label{slope}
  p \approxeq \frac{1}{3}L\rho_0,
\end{equation}
where $\rho_0$ is the nuclear saturation density. We can clearly observe a correlation in the vicinity of the nuclear saturation density. These EoS can be used in Eq. \eqref{tov} and one can obtain the corresponding masses and
radii. For Fig. \ref{mlall} we considered the maximum mass that each EoS can reach. These stars have an associated $\rho_c$ (or
$p_c$) that unlike $\rho_0$ cannot be reached on Earth, and the only way to determine stellar values is through an indirect method, i.e., through the maximum mass that one needs to compare with the observational data. No simple and direct
correlation is seen in Fig. \ref{mlall}, as it was clear in the case of Fig. \ref{p2L},
for what one needs machine learning techniques that are robust to changes in complex correlations and the emergence of structures in global properties. Here, we are
going to use a classification scheme, i.e., clustering techniques that will help us to make new predictions in future studies.

\section{Clustering}\label{sec:cluster}
Clustering methods were developed to study complex datasets and determine patterns and correlations among the degrees of freedom of the system (features). The idea is to
separate the elements into groups that share similarities, often too complicated to be identified at first glance. The method of clustering
falls into the category of {\it unsupervised learning} techniques \cite{geron/2017}.

\begin{figure}[h]
\centering
\includegraphics[scale=0.52]{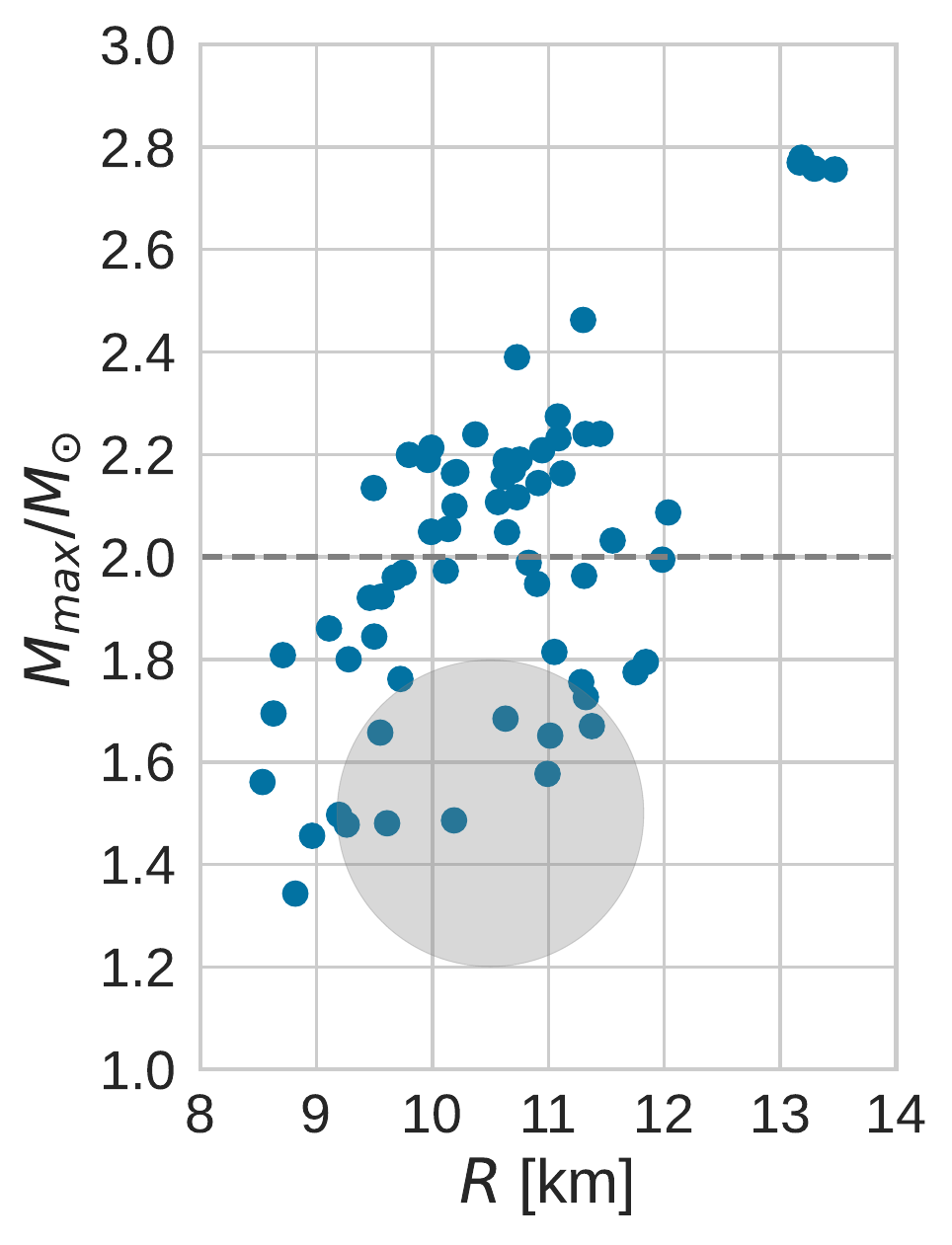}
\caption{\label{Set_of_EoS_from_LIGO}Maximum mass and its respective radius for 65 EoS from the LIGO {\it Lalsuite} \cite{lalsuite} library. The gray shaded circle shows the mass-radius constraints from GW170817 event. The horizontal dotted line at $2\ M_{\odot}$ is to remind the two massive pulsars J0348+0432 and J1614-2230 observed.}
\end{figure}

To compute correlations, find global structures, and clusters, we consider a larger set of EoS than our previous work \cite{lobato/2022}, where we were interested in the microscopic aspects. For this work, we use the 65 EoS from the LIGO {\it Lalsuite} \cite{lalsuite} library. In Fig. \ref{Set_of_EoS_from_LIGO} we show the maximum mass for all the EoS. We also show the possible region for the
mass-radius according to the gravitational wave observation
\cite{ligo/2018, ligo/2019} in a gray shaded circle. The horizontal dotted line marks two solar masses, since some pulsars above two solar masses have being observed \cite{demorest/2010, antoniadis/2013, linares/2018, cromartie/2020}.

\subsection*{$k$-means}
Here we employed the $k$-means clustering method \cite{geron/2017} to determine cluster structures present in the results shown in Fig.\ref{Set_of_EoS_from_LIGO}. The $k$ is the number of $k$-means clusters present in the data set. This is a method of quantization
\cite{linde/1980,geron/2017}, where it assumes that the clusters are defined by the distance of the points to their class centers only. The goal of clustering is to find the $k$-mean
vectors $c_1,\ldots,c_k$ and provide the cluster assignment $y_i\in \{1,...,k\}$ of each point $x_i$ in the set. The following criterion is used:
\begin{equation}
{\rm J} = \sum_{j=1}^k \sum_{i=j}^n
\|x_i-c_j\|^2 =\underset{\mathbf{S}}{\arg \min } \sum_{i=1}^K\left|S_{i}\right| \operatorname{Var} S_{i},
\label{eq:kmeans}
\end{equation}
where centroids are given by: $c_j = \frac{1}{m_j}\sum_{i=1}^{m_j}x_i$.

In this way, given the initial set, the algorithm goes between two steps: {\it (i) Assignment step}: Assign one element to the cluster with the nearest mean with the least squared Euclidean distance; {\it (ii) Update step}:
 Recalculate the centroids for the elements of each cluster. The $k$-means algorithm is based on an interleaving approach where the cluster assignments $y_i$ are established given the centers and the centers are then computed given the assignments.

\begin{figure}[h]
\centering
\includegraphics[scale=0.52]{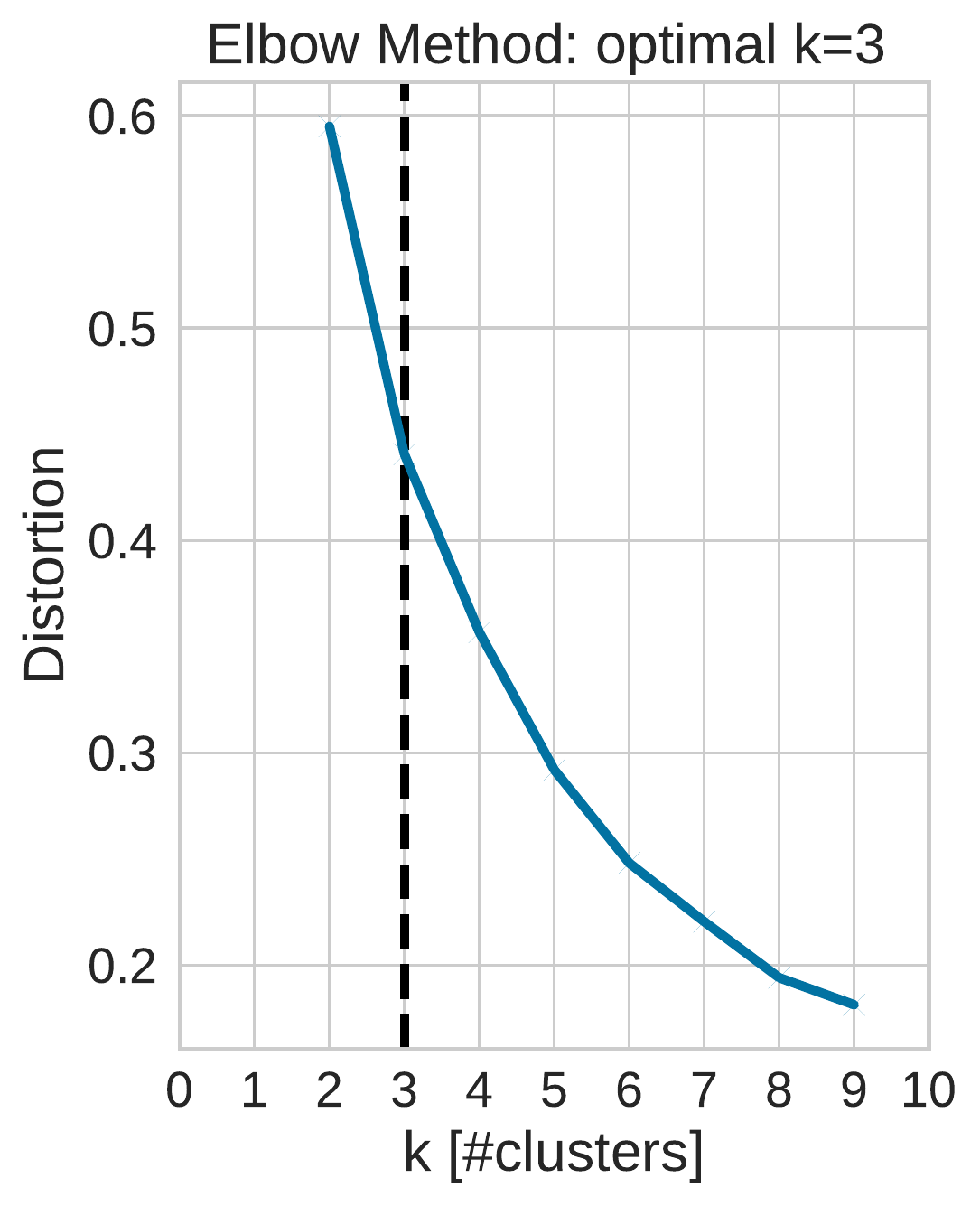}
\includegraphics[scale=0.52]{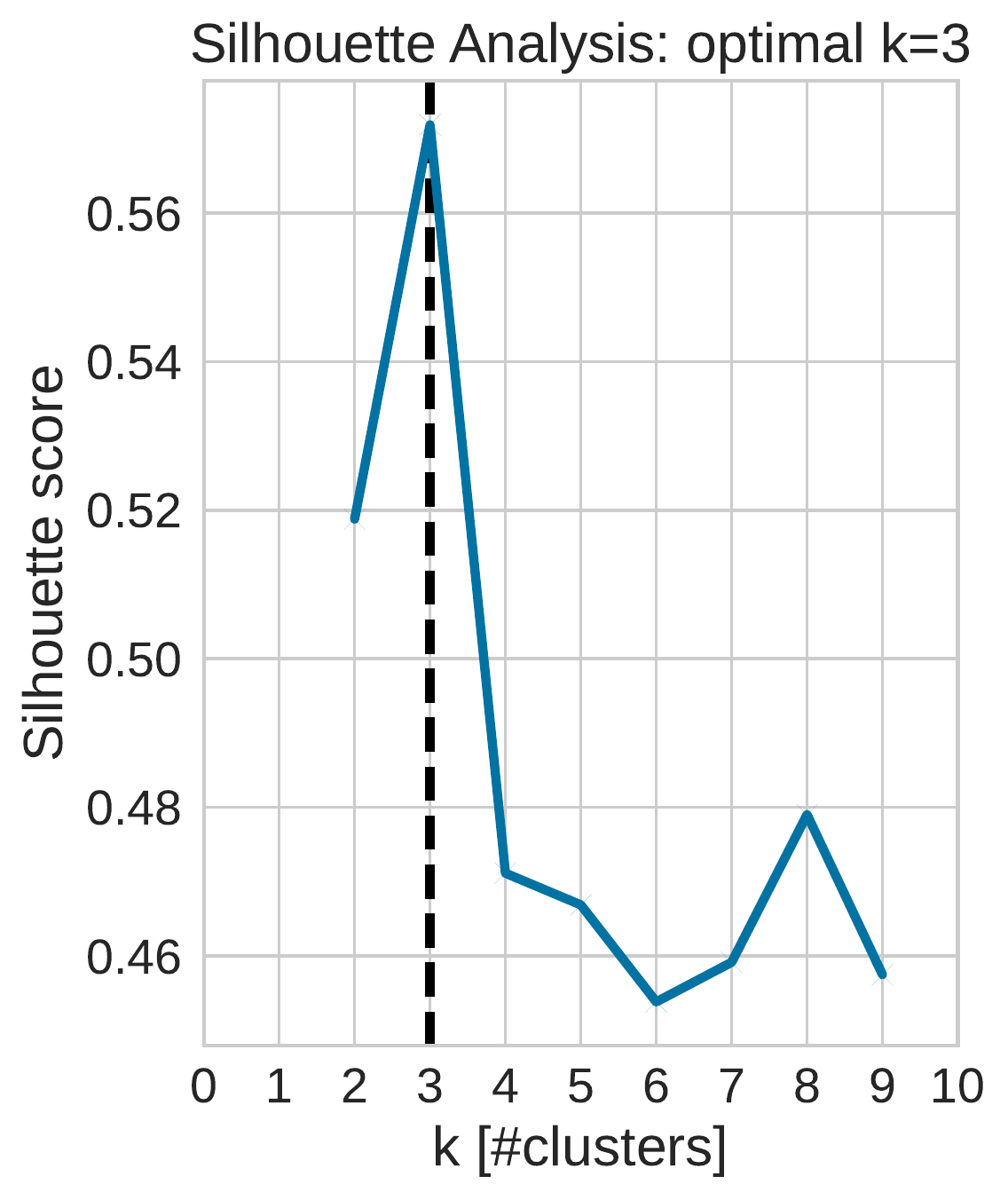}
\caption{\label{optimal}The Elbow and Silhouette methods to determine the optimal number of $k=3$ clusters.}
\end{figure}

The number of $k$ clusters is determined with the help of well-known procedures \cite{halkidi/2001,geron/2017}. We have used two different ways to arrive and confirm the value of three clusters ($k=3$). The first one is the {\it elbow point}, which determines the $k$ and the
homogeneity of the groups. On the left side of Fig. \ref{optimal} we plot the within-group homogeneity as a function of $k$. The ``elbow
point'' (vertical dashed line) represents the optimal for $k=3$ clusters. On the right side of the same Fig.\ref{optimal} we have the
silhouette plot, which measures the similarity/cohesion of the
elements to its clusters relative to other clusters employing a
distance metric map. The optimal value here is found at the highest inflection point that takes place at $k=3$, which confirms the number of three cluster present in our data set. More details about methods employed here can be found in \cite{geron/2017}.

\begin{figure}[h]
\centering
\includegraphics[scale=0.52]{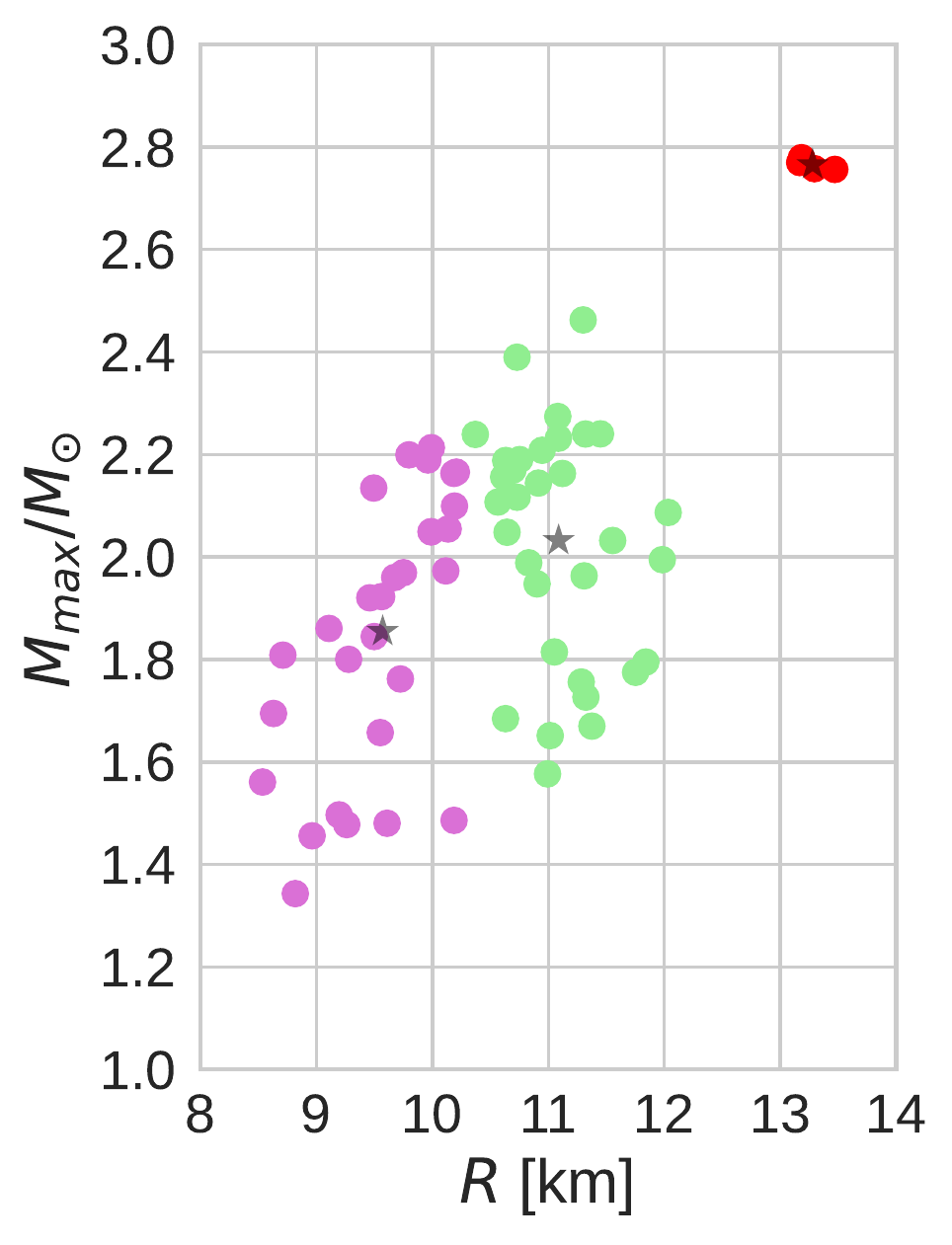}
\includegraphics[scale=0.52]{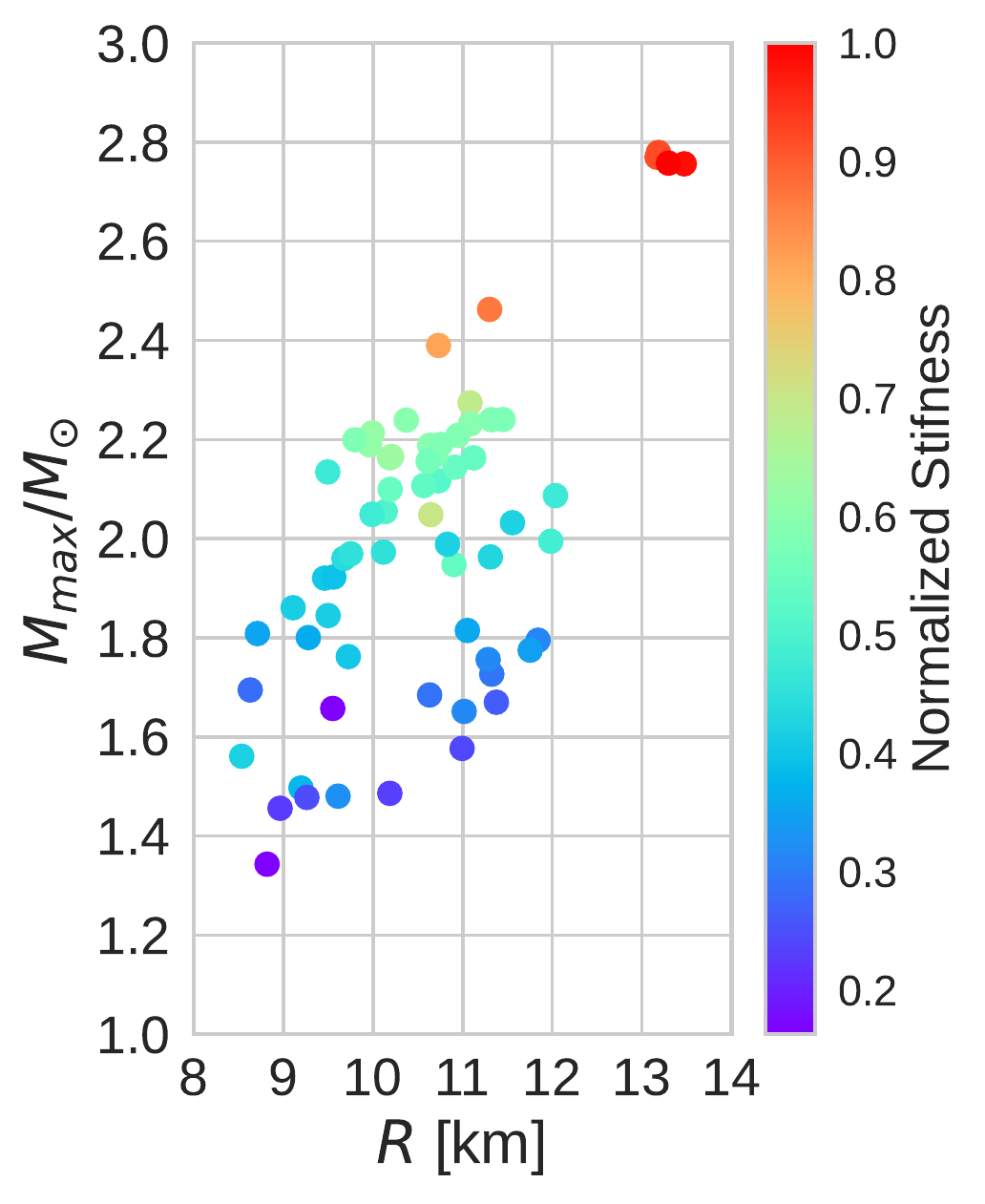}
\caption{\label{kmeans}Left side: Three clusters obtained in the maximum mass and its
  respective radius space utilizing $k$-means. Right side: Mass-radius distribution according to the normalized stiffness of the EoS.}
\end{figure}

In Fig. \ref{kmeans} (left panel), we show the resulting $k=3$ cluster structure obtained with the $k$-means approach. The three groups determined in this dataset are highlighted in different colors: the first one is pink, the second one in green
and finally the third one in red. The dark stars give the position of each cluster centroid. Each cluster contains many EoS with different microphysical characteristics. The separation into
groups represents similarities in traits or characteristics in this space $W$. On the right side of the same figure, we have the normalized stiffness of
the EoS, we have the increment in the mass that goes from the softness to the
stiff ones, the stiff EoS being responsible for the
most massive stars. The clusters depicted by the $k$-means approach can be associated to three regimes of the stiffness space, the high (red), intermediate (green) and low (purple) regimes.

\section{Concluding remarks}\label{sec:conc}
In this work, we have studied properties of a set of different EoS. We have solved the hydrostatic
equilibrium equations to obtain the maximum mass and radius for
each EoS. The outputs were used to define a vector space named $W$. The elements of this vector are taken to represent global characteristics of the microphysics of the equation of state for each model. Utilizing the $k$-means algorithm we were able to find structures in this space, leading to three subspaces, i.e.
three clusters. The structures found represent similarities in
traits/characteristics within the group only based in the elements of
$W$. The number of clusters $k=3$ was determined with two different
approaches frequently employed in ML studies. The resulting structures
can be associated to differences in the physics models used to derive
each EoS. We have shown that the clusters found by the unsupervised ML
approach can be associated to different stiffness of the equation of state. We believe that this approach can be combined with astronomical data to find additional correlations. This would help for a better understanding of extrapolation limits of current models.

This work is a
complementary study to our previous work \cite{lobato/2022}, where we
have analyzed the microphysics of the EoS. Combining the two works we
can also study the $V\otimes W$ space which will certainly bring new
correlations and bring light to the path of the correct EoS of neutron stars. We recall that the study of the whole M-R relation points is also important
when comparing with precise data from LIGO-VIRGO-KAGRA and NICER and
other astronomical data.

\section*{Acknowledgements}
RVL and CAB have been supported in part by the U.S. DOE Grant
No. DE-FG02-08ER41533. RVL also have been supported by Uniandes
University. This work is performed in part under the auspices of the U.S. Department of Energy by Lawrence Livermore National Laboratory under Contract DE-AC52-07NA27344. We would like to thank CNPq and INCT-FNA for paper publishing financial support.

\section*{References}
\bibliographystyle{iopart-num}
\bibliography{lib.bib}

\end{document}